\documentclass[prb,twocolumn,showpacs,floatfix,amsmath,amssymb,superscriptaddress]{revtex4-1}
\usepackage{amsfonts}
\usepackage{stmaryrd}
\usepackage{bbm}
\usepackage{mathrsfs}
\usepackage{tipa}
\usepackage{amssymb}
\usepackage{txfonts}
\usepackage{graphicx}
\usepackage{dcolumn}
\usepackage{epstopdf}
\usepackage[colorlinks,linkcolor=blue,urlcolor=blue,citecolor=blue]{hyperref}
\usepackage{multirow}
\usepackage{subfigure}
\usepackage{url}
\usepackage{upgreek}

\begin{document}

\newcommand*{\cm}{cm$^{-1}$\,}
\newcommand*{\Tc}{T$_c$\,}


\title{Field induced magnon excitation and in gap absorption of Kitaev candidate RuCl$_3$}
\author{L. Y. Shi}
\affiliation{International Center for Quantum Materials, School of Physics, Peking University, Beijing 100871, China}

\author{Y. Q. Liu}
\affiliation{International Center for Quantum Materials, School of Physics, Peking University, Beijing 100871, China}

\author{T. Lin}
\affiliation{International Center for Quantum Materials, School of Physics, Peking University, Beijing 100871, China}

\author{M. Y. Zhang}
\affiliation{International Center for Quantum Materials, School of Physics, Peking University, Beijing 100871, China}

\author{S. J. Zhang}
\affiliation{International Center for Quantum Materials, School of Physics, Peking University, Beijing 100871, China}

\author{L. Wang}
\affiliation{Institute of physics,Chinese academy of Sciences,Beijing, China}

\author{Y. G. Shi}
\affiliation{Institute of physics,Chinese academy of Sciences,Beijing, China}

\author{T. Dong}
\email{taodong@pku.edu.cn}
\affiliation{International Center for Quantum Materials, School of Physics, Peking University, Beijing 100871, China}

\author{N. L. Wang}
\email{nlwang@pku.edu.cn}
\affiliation{International Center for Quantum Materials, School of Physics, Peking University, Beijing 100871, China}
\affiliation{Collaborative Innovation Center of Quantum Matter, Beijing, China}


\begin{abstract}
We use time-domain terahertz spectroscopy to measure the low energy conductivity and magnons in RuCl$_3$ under external magnetic field. At zero field, an oscillation with a frequency of 0.62 THz is clearly observed in time-domain spectrum below T$_N$, which is identified as a magnon excitation in the magnetic order state. The magnon excitation is not affected by the external magnetic field $\textbf{H}_{DC}$ when it is applied along the c-axis, but is clearly suppressed when $\textbf{H}_{DC}$  is applied within ab plane. More interestingly, when the magnetic component of THz wave $\textbf{h}(t)$ is perpendicular to the applied in-plane magnetic field, we observe another coherent oscillation at slightly higher energy scale at the field above 2 T, which is eventually suppressed for $H_{DC}>$5 T. The measurement seems to indicate that the in-plane magnetic field can lift the degeneracy of two branches of low energy magnons at $\Gamma$ point. The low energy optical conductivity calculated from the measured transmission spectrum is dominated by a broad continuum contribution, which is not affected by changing either temperature or external magnetic field. The continuum is likely to be related to the fractional spin excitation due to dominated Kitaev interaction in the material.

\end{abstract}



\maketitle
\section{Introduction}
Spin orbit coupling (SOC) can give rise to highly nontrivial and exciting physics in quantum materials\cite{doi:10.1146/annurev-conmatphys-031115-011319}. For non-interaction systems, SOC can cause band inversion which is a key condition for realizing intriguing topological insulators\cite{RevModPhys.82.3045,  RevModPhys.83.1057}. When SOC meets with Hubburd U in correlated electronic materials, more striking physics could be induced cooperatively. For example, a combined strong SOC and crystal electric field effect can reorganize the transition metal 4d or 5d orbitals into J-multiplet, where J is the total angular momentum from the combined spin and orbital components, then a relatively week U is sufficient to localize single occupied J = 1/2 doublet, giving rise to a Mott insulator. Jackeli and Khaliullin showed that the magnetic interactions in such spin-orbit coupled Mott insulators are highly anisotropic and dictated by lattice geometry. A Kitaev model with bond-dependent interactions could be realized in such Mott insulator on a honeycomb lattice \cite{PhysRevLett.102.017205}. Since Ketaev model is a representative example of realizing quantum spin liquid (QSL), an exotic state of matter without magnetic long range order even at the lowest temperature, which can host fractional fermionic spin excitation (itinerant Majorana fermions and localized $Z_2$ fluxes) \cite{KITAEV20032}, it has attracted tremendous attention in condensed matter community.

Realization of the Kitaev model and physics in real materials has been pursued mainly in Ir-based honeycomb compounds, such as A$_2$IrO$_3$, A=Na or Li, where due to the strong SOC the multiorbital 5d $t_{2g}$ state can be mapped into a single orbital state with pseudospin $J_{eff}=1/2$ on every Ir-O octahedron with highly anisotropic nearest neighbor coupling. However, significant lattice distortions exist in the iridates that may lift $t_{2g}$ degeneracy and invalidate the $J_{eff}$ description. Recently, Mott insulator $\alpha$-RuCl$_3$  (hereafter RuCl$_3$) consisting of edge-sharing RuCl$_6$ octahedron in honeycomb lattice emerges as another candidate material \cite{PhysRevB.90.041112, PhysRevB.91.241110}. In contrast to the iridates, the RuCl$_6$ octahedron is much closer to cubic. In this material, $Ru^{3+}$ has an effective spin one half due to strong SOC and crystal field splitting. The J = 1/2 and J = 3/2 splitting in RuCl$_3$ has been determined by neutron scattering and Raman scattering to be about 195 meV \cite{PhysRevB.93.075144}.

Though RuCl$_3$ is a promising material for realizing Kitaev quantum spin liquid physics, it actually has a collinear "zigzag" magnetic order below 7.5 K (as displayed in Fig. \ref{Fig:1} (a) where domains with three different kinds of magnetic orientation are also shown), indicating presence of significant non-Kitaev interactions, such as Heisenberg or off dialog interactions \cite{Winter2017}. Such interactions generally lift the degeneracy of pure Kitaev ground state and introduce diverse spin solid states (stripy, zigzag), depending on the relative strength of interactions. For RuCl$_3$, the low temperature colinear zigzag-ordered magnetic structure (ferromagnetc zigzag chains, antiferrmagnetic coupled to each other) with a ordered  moment of $0.4\pm 0.1 \mu_B /Cu^{2+}$ is one of the magnetic states predicted by the Kitaev-Heisenberg or extended models \cite{RN100,Winter2017}. The relatively low ordering temperature and moment reduction of $35\%$ suggest that, despite the presence of inter-layer coupling, RuCl$_3$ exhibits strong quantum fluctuations and is in proximity to Kitaev spin liquid. Even above the ordering temperature, properties proximate to the Kitaev spin liquid were reported in the recent experiments such as Raman \cite{PhysRevLett.114.147201,  RN37} and inelastic neutron scattering measurements \cite{Banerjee1055,  PhysRevLett.118.107203,  RN133}. A theoretical study also indicates that the spin-liquid ground state survives even when a small Heisenberg interaction is added to the Kitaev model\cite{PhysRevLett.105.027204}. On the other hand, a moderate in plane magnetic field of 7.5 T is sufficient to completely suppress the long range magnetic order, inducing a quantum spin liquid state\cite{PhysRevLett.119.037201}. However, the nature of field induced spin liquid state and the character of low energy spin excitations remain to be clarified. Earlier thermal conductivity \cite{PhysRevLett.118.187203} and nuclear magnetic measurement (NMR)\cite{PhysRevLett.119.227208} in high magnetic field suggested presence of gapless excitations in field induced spin disordered state. On the contrary, other experiments reveal a clear spin gap in excitation spectrum\cite{PhysRevB.95.180411,  PhysRevLett.119.037201,  PhysRevB.96.041405}. A more recent thermal conductivity measurement under magnetic field indicate no sign of gapless magnetic excitation\cite{PhysRevLett.120.067202}.

 In this report, we investigate the low energy spectrum of $RuCl_{3}$ due to magnetic excitation by terahertz (THz) time domain spectroscopy. Although there exist a few THz spectroscopy measurements \cite{PhysRevLett.119.227201,PhysRevLett.119.227202}, the present work represents a more comprehensive study where the magnetic component of probing terahertz relative to the external magnetic field is systematically varied.  We observe new collective excitation and phenomena and discuss their possible implications.

\section{Experiments}
Single crystal of RuCl$_3$  were grown from commercial RuCl$_3$ power by vacuum sublimation in sealed quartz tubes. This result in flat, plate like crystals with typical size $5\times3\times0.5 mm^3$, as shown in Fig.\ref{Fig:1} (b). Magnetization and heat capacity were measured as function of temperature via physical property measurement system with fields up to 16 T. The heat capacity measurements were done with a single crystal mounted on a vertical sample holder where the magnetic field was applied parallel to honeycomb plane. Compared with previous transport measurements, our sample undergoes a single antiferromagnetic transition at $T_N$ =7.5 K, as illustrated in Fig1 (c). As the multilple transitions are linked to stacking faults of the layer coupling\cite{PhysRevB.92.235119}, a single phase transition implies a nearly uniform stacking pattern of the sample.

Time domian terahertz spectra (TDTS) of $RuCl_{3}$ were measured using home building spectroscopy equipped helium cryostat and Oxford spectramag (magnetic field up to 7 T). The probing terahertz radiation was generated by large area GaAs photoconductive antenna and detected by 1 mm ZnTe crystal via free space electro-optics sampling method\cite{doi:10.1063/1.116356}. The femtosecond laser delivered from the mode locked Ti:sappire (center wavelenght of 800 nm, pulse duration of 35 fs, and 80 MHz repetition rate) was used as pumping source. The linear polarized terahertz light was focused on the sample by $90^{\circ}$ off axis parabolic mirrors. The setup was placed within the box filled with dry nitrogen to eliminate THz absorption by water in the atmosphere. In a typical TDTS experiment, the electric field of a THz pulse passing through a sample, and the same size aperture as a reference is recorded as a function of delay time. Fourier transformation the recorded time traces provides the frequency dependent complex transmission spectra including magnitude and phase information. For transparent compounds, complex transmission formula derived in normal incidence by Fresnel model is given by:
\begin{equation}
\widetilde{T}=\frac{4\widetilde{n}}{{(\widetilde{n}+1)^2}}\exp(\frac{i\omega d}{c}(\widetilde{n}-1))
\end{equation}
where $\widetilde{n}$ is the complex refractive index of sample, $\omega$ the frequency, d the sample thickness, and $c$ the speed of light. Numerical itinerate method was used to solve this complex equation to get both the frequency dependent real and imaginary parts of optical constants without Kramers-Kroning transformation\cite{terahertz}.

The superconducting magnet was mounted on a rotating stage, in order to perform measurement with magnetic field in Faraday (light $\textbf{k}\parallel \textbf{H}_{DC}$) and Voigt geometry (light $\textbf{k}\perp \textbf{H}_{DC}$) configuration. The orientation of magnetic field supplied from the superconducting  magnet can rotate from 0 to 90 degree relative to terahertz propagating direction conveniently. The magnetic dependent measurement were mainly performed at 2 K.

\section{magnon excitation }

\begin{figure}
\includegraphics[width=8.5cm]{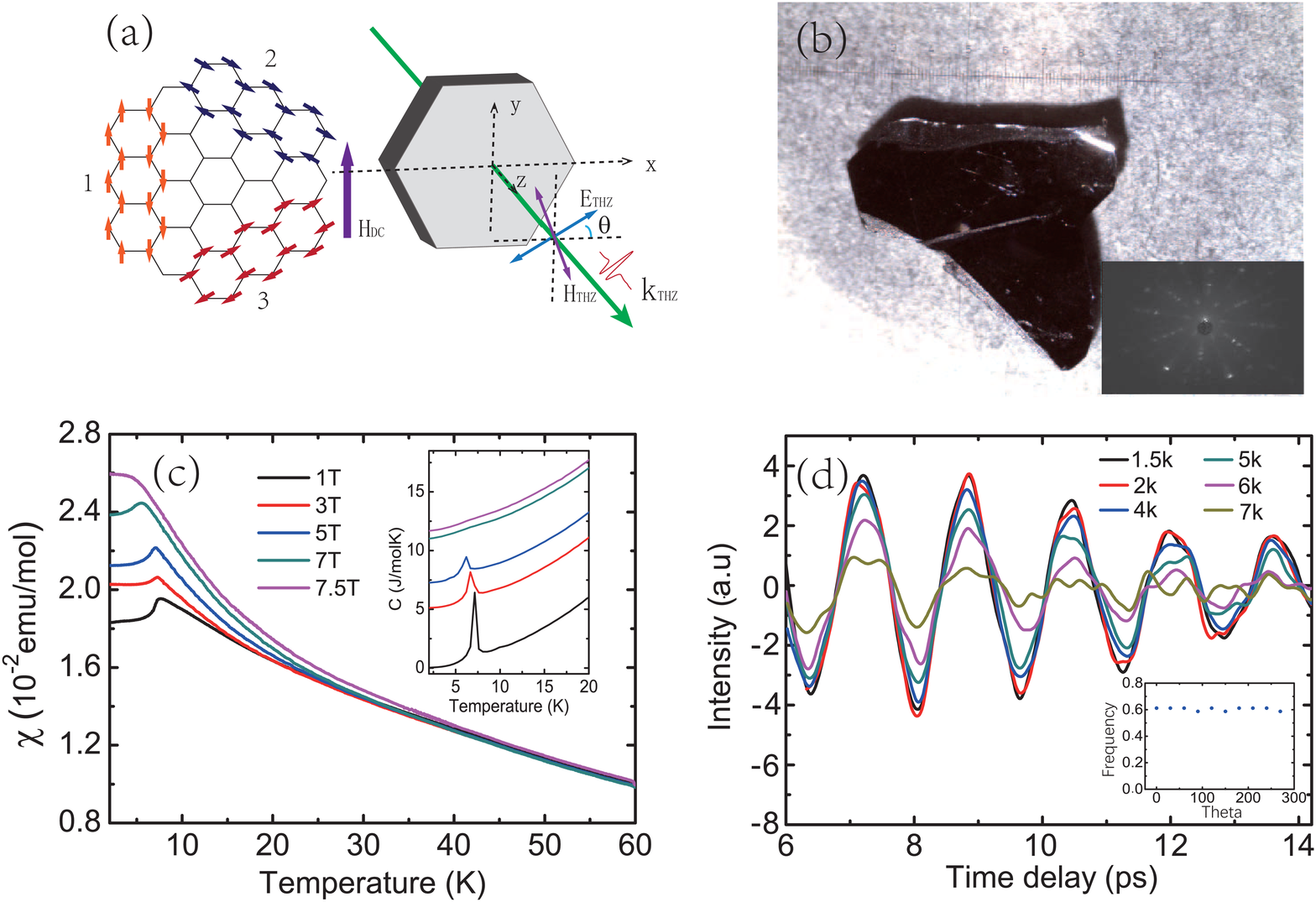}
\caption{\label{fig:epsart}(a) Schematic of three domain orientations with zigzag magnetic order and terahertz measurement. (b) Picture of our sample. (c) Magnetic suscepbility of RuCl$_3$ at selected fields, The inset presents the field dependent heat capacity. (d) Coherent magnon excitation in time domain at selected temperatures, subtracting 8 K time traces at same magnetic field. The inset shows angle dependence of magnon excitation at 4 K.}\label{Fig:1}
\end{figure}

\begin{figure*}
\includegraphics[width=17cm]{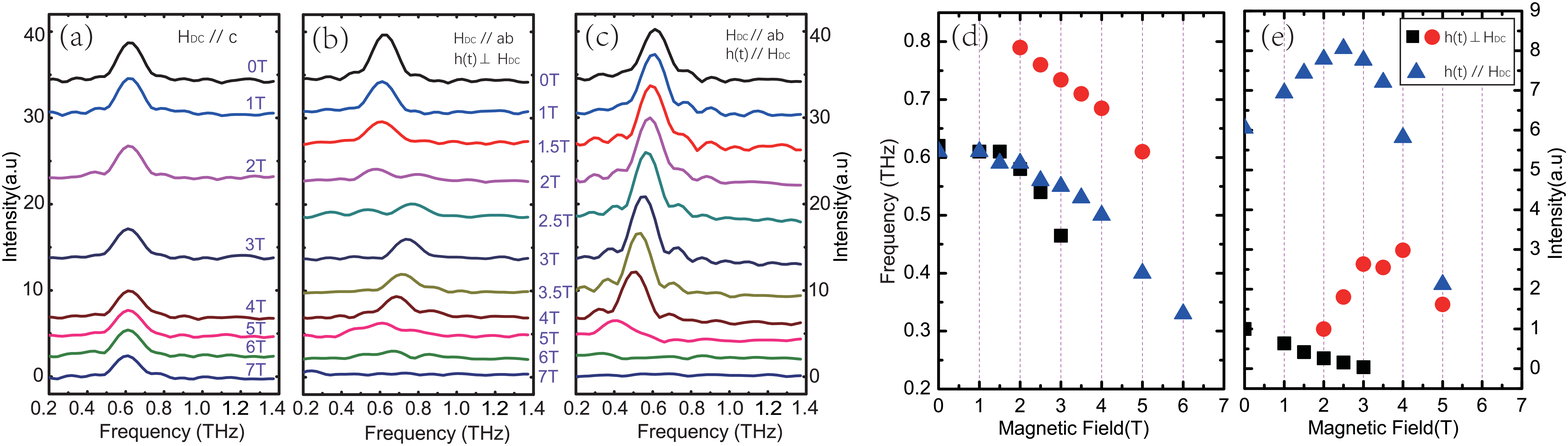}
\caption{\label{fig:epsart}Coherent magnon excitation in time domain at selected magnetic fields, subtracting 8K time traces at same magnetic field. (a) Spectra of the magnon measured at selected magnetic field with $\textbf{H}_{DC}\parallel c$ configuration. The magnon excitation located at Brillioun zone center does not change with magnertic field. (b) Magnon excitation measured at selected magnetic fields with $\textbf{H}_{DC}\parallel ab$ and $\textbf{h}(t)\perp\textbf{H}_{DC}$  configuration. (c) Magnon excitation measured at selected magnetic fields with $\textbf{H}_{DC}\parallel ab$ and $\textbf{h}(t)\parallel\textbf{H}_{DC}$ configuration. (d) and(e)The evolution of magnon intensity and frequency with magnetic field. }\label{Fig:2}
\end{figure*}

Figure \ref{Fig:1} (d) shows time traces for various temperatures below $T_N$, where all traces were subtracted by THz time trace measured at 8 K. As can be seen, the clear oscillations are present below $T_N$ arising from coherent spin procession induced by magnetic component of terahertz wave. In zero magnetic field, a signature of magnon $\Omega_1$ located at 0.62 THz is discerned when temperature is below 8 K, in accordance with previous time domain terahertz spectroscopy \cite{PhysRevLett.119.227201,PhysRevLett.119.227202}, electron spin resonance (ESR)\cite{PhysRevB.96.241107,Wellm2017} and neutron scattering measurement in zero magnetic field \cite{Banerjee1055}. From angle dependent measurement, we find that magnon is insensitive to orientation of THz radiation relative to zigzag chain. As shown in the inset of Fig.\ref{Fig:1}(d), the frequency of magnon excitation is nearly isotropic. The long wavelength spin wave excitations are well defined transverse modes, i.e., they involve oscillations of the magnetism in a direction perpendicular to the ordered moments, $\triangle \textbf{S}\perp\textbf{S }$. In general, the magnetic component of THz wave couples with the magnetic moment of sample through Zeeman torque $dM/dt=\gamma$ \textbf{M} $\times$ \textbf{$H_{THz}$}, where $\gamma $ denotes the gyromagnetic constant, \textbf{M} and \textbf{$H_{THz}$} are the macroscopic magnetization and the impulsive magnetic field of probing THz wave respectively\cite{PhysRevLett.110.137204}. Because the wavelength of terahertz light greatly exceeds atomic length scale, the spin wave excitation measured by terahertz spectroscopy locates at the zone center. One can speculate that when the magnetic component of terahertz radiation is parallel to the orientation of magnetic moment, the transverse modes should be silent. However, in our measurements, the spin wave excitations are nearly isotropic. This finding can be explained as a result of existence of magnetic domains in $RuCl_{3}$\cite{PhysRevB.95.180411}. Typically, three possible zigzag domains will form within the sample as it is cooled below $T_N$, as presented in Fig. \ref{Fig:1} (a). When we rotate the orientation of probing terahertz radiation, different magnetic domains will contribute to spin precession, giving rise to nearly isotropic magnon response. As will be shown below, this scenario also explain the magnetic field dependent measurement very well.

When we apply magnetic field parallel to the c-axis (\textbf{H}$_{DC}$ $\parallel$ c) at 2 K  and measure in plane optical response (Faraday geometry), we did not observe any systematic dependence of the magnon oscillations as shown in Fig.\ref{Fig:2}(a). In particular, the magnon excitation is essentially unchanged with the magnetic field. The slight change of the magnitude of magnon oscillations  is within our experimental uncertainly, including the possible clipping of the terahertz light on the magnet inner optical window  and slight distortion of the optics under magnetic field. This result is consistent with our heat capacity measurements, where heat capacity jump persist up to 8 T when magnetic field perpendicular to ab plane (not shown).

The rotation of external magnetic field from the c axis to the ab plane (\textbf{H}$_{DC}$ $\parallel$ ab) affects the spin excitation modes significantly. For $\textbf{h}(t)$ $\perp$ $\textbf{H}_{DC}$ configuration, an additional, magnetic field induced higher frequency mode $\Omega_2$ ($\sim$0.7-0.8 THz) appears. The intensity of this mode increases slightly for $2T<H<4T$, then drops gradually, as illustrated in Fig.\ref{Fig:2}(b). In contrast, the intensity of lower mode $\Omega_1$ decreases monotonously with increasing magnetic field. In a narrow range of magnetic field 2.0-2.5 T, the low frequency mode $\Omega_1$ and higher excitation $\Omega_2$ coexist. To identify the selection rules of observed two collective excitations, we rotate the polarization of terahertz radiation, making magnetic component of terahertz radiation parallel to external magnetic field. In this configuration, we can not resolve any additional higher frequency mode as shown in Fig.\ref{Fig:2}(c), which is consistent with previous measurements \cite{PhysRevLett.119.227201,  PhysRevLett.119.227202}. In the \textbf{h}(t) $\parallel$ $\textbf {H}_{DC}$ configuration, the frequency of $\Omega_1$ shows red-shift gradually with magnetic filed increasing, and the intensity of the peak increases firstly when magnetic field is lower than 2.5 T, and begins to lose intensity above this field. The main result of the magnetic field dependence of the intensity and frequency is further summarized in Fig.\ref{Fig:2}(d) (e). While the lower energy mode $\Omega_1$ persists up to 6 T in $\textbf{h}(t) \parallel \textbf{H}_{DC}$ configuration, such modes lose its intensity at 3.5 T completely in $\textbf{h}(t) \bot\textbf {H}_{DC}$  configuration. In the following, we shall argue that this intensity evolution versus magnetic field can be explained by domain reorientation.

As illustrated in Fig.\ref{Fig:1} (a), the magnetic field $\textbf{H}_{DC}$ is parallel to zigzag direction of domain 1 (i.e. b-axis) in all of our measurements. From previous neutron scattering measurements under magnetic field, the domain reorientation with a spin flop-like effect occurs gradually with field and reaches equilibrium above 2 T\cite{PhysRevB.95.180411}. As a result, spin moments orientation are nearly perpendicular to the external magnetic field $\textbf{H}_{DC}$ above 2 T. In $\textbf{h}(t)\parallel\textbf{H}_{DC}$ configuration, magnon excitation arising from domain 1 should be silent at zero magnetic field. Then applied magnetic field redistributes domain population from domain 1 to domain 2 and 3, which induces an intensity increase of magnon excitation at low field. The signal is suppressed and eventually vanishes at higher magnetic field due to the suppression of antiferromagnetic order. For $\textbf{h}(t)\bot\textbf{H}_{DC}$ geometry, at zero magnetic field, all of the domains are expected to be optically active. In partiucular, the magnon excitation from domain 1 should have the highest signal since their moments are perpendicular to the terahertz probe $\textbf{h}(t)$.  As the magnetic field rotates the orientation of moments in domain 1, the signal level from domain 1 should be gradually reduced, leading to a reduction of overall signal. This is what we observed in Fig.\ref{Fig:2} (b) below 2 T.

Having interpreted the evolution of $\Omega_1$ mode, we now focus on the origin of the new $\Omega_2$ mode near 0.7-0.8 THz at higher field. Notably, a similar mode was observed in magnetic field dependence ESR measurement, in which such mode was assigned to an exchange mode at Brillouin zone boundary whose optical transition (\textbf{q}=0) is normally forbidden but become allowed with the assistance of Dzayaloshinskii-Moriya (DM) interaction\cite{PhysRevB.96.241107}. It should be noted that the $\Omega_2$ mode only can be detected in the $\textbf{h}(t)\bot\textbf{H}_{DC}$ configuration with $H_{DC}>2 T$. If the DM interaction assisted picture is valid, the magnetic field dependent selection rules needs to be carefully examined theoretically. An alternative possibility is that the $\Omega_2$ mode is also a magnon at $\Gamma$ point. Theoretically, it is known that there exist 4 branches of spin wave dispersions for the zigzag magnetic order on the honeycomb lattice arising from the presence of Heisenberg or off-diagonal coupling besides the Kitaev interaction. A calculation based on the linear spin wave theory indicates that the two lowest ones, which are degenerate at $\Gamma$ point in the Brillouin zone, have the energy very close to the experimentally observed energy near 3 meV \cite{Winter2017}. It is likely that the external in-plane magnetic field can lift the degeneracy of the two branches of magnons at $\Gamma$ point, therefore leading to the observation of the new mode by THz and ESR measurement. It also deserves to remark that the neutron scattering measurements from different groups show very different spin wave dispersions near $\Gamma$ point \cite{Banerjee1055,PhysRevLett.118.107203,RN133}. Further explorations are still needed to solve the discrepancy.

\section{low energy optical conductivity }

 In addition to detection of magnon excitations, low energy optical conductivity also yields useful information about continuum exciations. In general, magnetic excitations are not directly electric active, but can become visible in the terahertz or infrared spectra through magneto-elastic and magnetic-electric effect. For U(1) gapless quantum spin liquid ground state, it was proposed earlier that spin degree of freedom can interact with charge through emerge gauge field, giving rise to a nontrivial absorption feature within charge gap, that is, a power law frequency dependence \cite{PhysRevLett.99.156402,  PhysRevLett.111.127401}. The extended Kitaev model that describes the zigzag ground state consisting a ferromagnetic Kitaev term and Heisenberg or off-diagonal terms pushes the RuCl$_3$ to the boundary of quantum spin liquid phase. The fact that the Kitaev term is considerably larger than other interactions implies that the degree of frustration is quite strong and should contribute to the low lying physics. A theoretical calculations indicate a linear-$\omega$ dependence of the spectral function at the center of Briluion zone\cite{PhysRevLett.117.037209}.

\begin{figure}
\includegraphics[width=8cm]{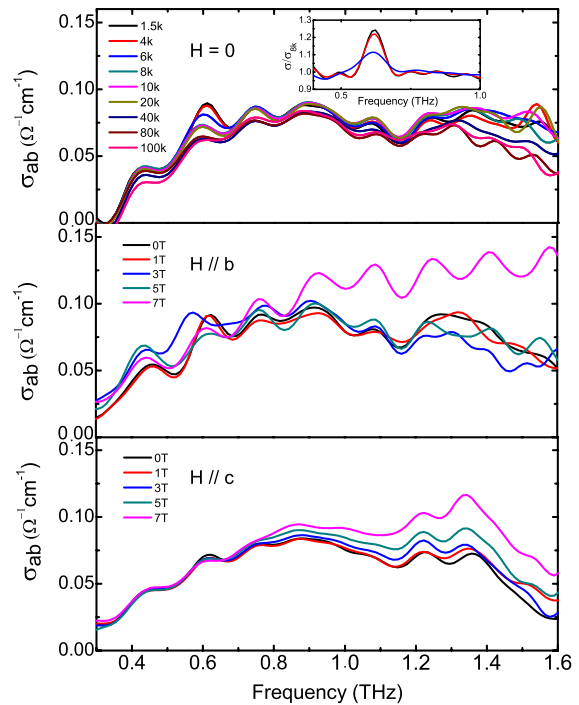}
\caption{\label{fig:epsart}In plane optical conductivity $\sigma_1(\omega)$ of $\alpha$-RuCl$_3$ measured at 0.3 THz-1.6 THz at selected temperature and magnetic field. (a)Temperature dependence of $\sigma_1(\omega)$  at zero magnetic field; (b) Magnetic field dependence of $\sigma_1(\omega)$ at 2 K with $\textbf{H}_{DC}\parallel ab-plane$ (in the b-axis); (c) Magnetic field dependence of $\sigma_1(\omega)$ at 2 K with $\textbf{H}_{DC}\parallel c-axis$.}\label{Fig:3}
\end{figure}

 Figure \ref{Fig:3}(a) presents real part of in-plane optical conductivity $\sigma_1(\omega)$ at different temperatures in the zero magnetic field. $\sigma_1(\omega)$  exhibits a finite broad peak centered around 0.8 THz at all measurement temperatures. Below $T_N$, the magnon absorption near 0.62 THz can be resolved on top of the background, which can be more clearly seen as a sharp peak in the ratio plot of the conductivity below $T_N$ over that at 8 K, as shown in the inset of Fig.\ref{Fig:3} (a). Compared with broad absorption, the spectral weight of pure magnon excitation is extremely small. Intriguingly, within our experimental accuracy, we can not resolve any temperature dependence of this broad absorption. Similar effect was also observed in reported THz measurement \cite{PhysRevLett.119.227201}. The results are very different from Raman and neutron scattering measurements on $RuCl_{3}$, where a continuum excitation due to fermionic excitations is found to increase with decreasing temperature only below 100 K \cite{PhysRevLett.114.147201}.

Figure \ref{Fig:3} (b) and (c) show the in-plane optical conductivity $\sigma_1(\omega)$ under magnetic field at a temperature of T=2 K with $\textbf{H}_{DC}\parallel c-axis$ and $\textbf{H}_{DC}\parallel ab-plane$ (in the b-axis), respectively. We did not observe any systematic dependence of the broad absorption on magnetic field. In contrast, the week peak contributed by spin wave excitation was suppressed by applied in plane magnetic field, as seen in Fig.\ref{Fig:3} (c). It is worth noting that at relatively high magnetic field ($\geq 6 T$), optical beam lines were slightly distorted by the high field, thus yielding artificial anomaly at high frequency. So, the separation of the curve at 7 T from others is not intrinsic. Since the continuum spectra are essentially not changed by increasing the temperature from 2 K to 300 K, it is not surprising that the broad spectra are not affected by applying a moderate magnetic field.

 The origin of this broad absorption is a crucial issue. Although the energy scale of charge gap in RuCl$_3$ is still under debate, the existence of an energy gap exceeding 200 meV is well established\cite{ROJAS1983349,  PhysRevB.96.165120,  PhysRevB.93.075144}. Accordingly, itinerant carriers in RuCl$_3$ are absent even at room temperature. Therefore, any possibility arising from charge excitation can be ruled out. In the charge gapped solids, charge-neutral bosonic collective mode can give rise to a finite resonant absorption within charge gap, for example, electric active magnon in multiferroics \cite{RN132}and phase mode in charge density wave insulators\cite{PhysRevB.66.024401}. However, there is no indication of multiferroic and collective charge excitations in RuCl$_3$. Therefore, quantum fluctuation from spin degree of freedom remains as the most possible origin. Particularly, absorption arising from fractional spin excitation due to dominated Kitaev interaction is an appealing interpretation for such continuum. This scenario is consistent with reduced magnetic moment, suppressed order temperature and presence of spin continuum excitation as revealed by neutron measurement\cite{RN133}. A recent theoretical calculation on Katiev materials by taking account of the complex interplay of the Hund's coupling, SOC and a trigonal crystal field distortion indeed showed that the fractionalized magnetic excitations, although emerging from an effective spin Hamiltonian, responded to an external AC electric field, which therefore explained the continuum THz response \cite{Bolens2017}. However, our experimental observation of the broad absorption being insensitive to either temperature (even up to room temperature) or magnetic field remains a challenge issue and needs further studies.

We notice that the magnitude of our measured optical conductivity has the same order as that of Herbertsmithite\cite{PhysRevLett.111.127401}, in which external electric field couples linearly to the emerge gauge field, inducing a power law behavior of low energy optical conductivity far below the charge energy gap\cite{PhysRevB.87.245106}. Nevertheless, for the Kitaev compound RuCl$_3$, the frequency dependence of low energy optical conductivity appears to be different. It exhibits a the broad peak centered around 0.8 THz. The results imply that different interactions involved in those quantum spin liquid candidates may lead to different low energy electrodynamic responses.

\section{Conclusion }
To conclude, we have investigated the low energy electrodynamic response of Kitaev candidate RuCl$_3$ by performing time domain terahertz spectroscopy measurement under magnetic field. A magnon excitation at the energy of 0.62 THz is clearly resolved below T$_N$. The excitation is not affected by an external magnetic field $\textbf{H}_{DC}$ applied along the c-axis but suppressed when applied within ab-plane. In particular, when the magnetic component of THz wave $\textbf{h}(t)$ is perpendicular to the applied in-plane magnetic field $\textbf{H}_{DC}$, we observe a new magnetic field induced mode at slightly higher energy scale, which is eventually suppressed for $H_{DC}>$5 T. The observation of two modes under magnetic field might be due to the lifting of the degeneracy of two branches of magnons at $\Gamma$ point by in-plane magnetic field. Besides the extremely small spectral weight by magnon, the low energy optical conductivity is dominated by a broad continuum contribution. We elaborate that this broad continuum contribution is likely to arise from the fractional spin excitation due to dominated Kitaev interaction in the material. However, its insensitivity to either temperature or magnetic field needs to be further explored.

\begin{acknowledgments}

We thank Yuan Li for helpful discussions. This work was supported by the National Science Foundation of China (No.11327806, No.11404385, GZ1123) and the National Key Research and Development Program of China (No.2016YFA0300902, 2017YFA0302904).
\end{acknowledgments}

\bibliographystyle{apsrev4-1}
\bibliography{RuCl3-terahertz}

\end{document}